\documentclass[apj,graphicx]{emulateapj}

\usepackage{graphicx, natbib}

\begin{document}

\shorttitle{Events leading up to the June 2015 outburst of V404 Cyg}
\shortauthors{Bernardini et al.}

\title{Events leading up to the June 2015 outburst of V404 Cyg}

\author{F.~Bernardini \altaffilmark{1,2}}
\author{D.M.~Russell\altaffilmark{1}}
\author{A.W.~Shaw\altaffilmark{3}}
\author{F.~Lewis\altaffilmark{4,5}}
\author{P.A.~Charles\altaffilmark{3,6}}
\author{K.I.I.~Koljonen\altaffilmark{1}}
\author{J.P.~Lasota\altaffilmark{7,8}}
\author{J.~Casares\altaffilmark{9,10,11}}

\email{bernardini@nyu.edu}

\affil{\altaffilmark{1}{New York University Abu Dhabi, P.O. Box 129188, Abu Dhabi, United Arab Emirates;bernardini@nyu.edu}}
\affil{\altaffilmark{2}{INAF $-$ Osservatorio Astronomico di Capodimonte, Salita Moiariello 16, I-80131 Napoli, Italy}}
\affil{\altaffilmark{3}{Department of Physics \& Astronomy, University of Southampton, Southampton, SO17 1BJ, UK}}
\affil{\altaffilmark{4}{Faulkes Telescope Project, School of Physics \& Astronomy, Cardiff University, The Parade, CF24 3AA, Cardiff, Wales}}
\affil{\altaffilmark{5}{Astrophysics Research Institute, Liverpool John Moores University, 146 Brownlow Hill, Liverpool L3 5RF, UK}}
\affil{\altaffilmark{6}{ACGC, University of Cape Town, Private Bag X3, Rondebosch, 7701, South Africa}}
\affil{\altaffilmark{7}{Institut d'Astrophysique de Paris, CNRS et Sorbonne Universit\'es, UPMC Paris~06, UMR 7095, 98bis Bd Arago, 75014 Paris, France}}
\affil{\altaffilmark{8}{Nicolaus Copernicus Astronomical Center, Bartycka 18, 00-716 Warsaw, Poland}}
\affil{\altaffilmark{9}{Instituto de Astrof\'isica de Canarias, E-38205 La Laguna, Santa Cruz de Tenerife, Spain}}
\affil{\altaffilmark{10}{Departamento de Astrof\'isica, Universidad de La Laguna, E-38206 La Laguna, Santa Cruz de Tenerife, Spain}}
\affil{\altaffilmark{11}{Department of Physics, Astrophysics, University of Oxford, Denys Wilkinson Building, Keble Road, Oxford OX1 3RH, UK}}

\begin{abstract}

On 2015 June 15 the burst alert telescope (BAT) on board {\em Swift} detected an X-ray outburst from the black hole transient V404 Cyg.
We monitored V404 Cyg for the last 10 years with the 2-m Faulkes Telescope North in three optical bands (V, R, and i$^{'}$). We found that, one week prior to this outburst, the optical flux was 0.1--0.3 mag brighter than the quiescent orbital modulation, implying an optical precursor to the X-ray outburst. There is also a hint of a gradual optical decay (years) followed by a rise lasting two months prior to the outburst.
We fortuitously obtained an optical spectrum of V404 Cyg 13 hours before the BAT trigger. This too was brighter ($\sim1\rm\,mag$) than quiescence, and showed spectral lines typical of an accretion disk, with characteristic absorption features of the donor being much weaker. No He II emission was detected, which would have been expected had the X-ray flux been substantially brightening. This, combined with the presence of intense H$\alpha$ emission, about 7 times the quiescent level, suggests that the disk entered the hot, outburst state before the X-ray outburst began. We propose that the outburst is produced by a viscous-thermal instability triggered close to the inner edge of a truncated disk. An X-ray delay of a week is consistent with the time needed to refill the inner region and hence move the inner edge of the disk inwards, allowing matter to reach the central BH, finally turning on the X-ray emission.

\end{abstract}


\section{Introduction}

In Low Mass X-ray Binaries (LMXBs) a black hole (BH) or a neutron star (NS) accretes matter from a low mass ($M\sim\,M_{\odot}$) Roche lobe filling companion. Many LMXBs are transient, alternating between long periods of quiescence (years), where the X-ray luminosity is faint ($\leq10^{33}\rm\,erg/s$) to shorter episodes of outburst where the X-ray luminosity strongly increases ($10^{37-38}\rm\,erg/s$) and can approach the Eddington limit. 

\cite{coriat12} showed that the global behaviour of LMXB outbursts is well described by the thermal--viscous disk instability model \citep[DIM; see e.g.][]{cannizzo93,lasota01}, where in quiescence a cold, non-stationary disk fills-up with matter until at some radius its temperature having reached a critical value triggers an outburst. Heating fronts propagate through the disk bringing it to a hot, quasi-stationary, bright state at which the X-ray luminosity reaches its maximum. The DIM can broadly explain the LMXB outburst cycle only if the inner disk is truncated during quiescence \citep[][hereafter DHL]{dubus01}.

The accretion flow structure inside the truncated disk's hole during quiescence forms a hot, optically thin, radiatively inefficient plasma \citep[see e.g.][]{narayan97,narayan08} that shines in X-rays, and/or a jet \citep{hynes09,xie14}, while the outer disk remains cool. However, from a DIM perspective the inner flow is not important, as long as it does not contribute to the dynamics of the disk (DHL).

V404 Cyg ($=$GS 2023+338), hereafter V404, is one of the closest and best studied LMXBs \citep[$2.39\pm0.14$ kpc;][]{millerjones09}. It hosts a $9\pm^{0.2}_{0.6}\,M_{\odot}$ BH \citep{khargharia10} with an orbital period of $6.4714\pm0.0001$ days \citep{casares94}.
V404 showed at least three outbursts \citep[1938, 1956, and 1989;][]{makino89,richter89,zycki99}, now followed by the burst alert telescope \citep[BAT;][]{barthelmy00} triggering on V404 on 2015 June 15 18:31:38 UT \citep[MJD 57188.772;][]{barthelmy15} when the first X-ray flare of a new outburst was detected, followed by multiple flares in hard X-rays with fast rise time and durations of hours \citep[][]{rodriguez15b}.

An optical precursor of the first X-ray flare was detected with the 2-m Faulkes Telescope North (FTN) a week before the first BAT trigger \citep{bernardini15}. 
Due to the sporadic nature of LMXB outburst start times, to the small number of transient sources currently known, and to the lack of regular high S/N monitoring (nowadays difficult at X-ray wavelengths but easier at optical wavelengths), it is hard to detect a delay in the rise to outburst between short (X-ray) and long (IR-optical) wavelengths. 
 
Only five other sources have shown an indication of similar behavior \citep{orosz97,shahbaz98,jain01,wren01,uemura02,buxton04,zurita06}.
This X-ray delay of uncertain origin, is reminiscent of the well-documented dwarf-nova outbursts 
UV to optical delay \citep[][and references therein]{smak98}. 
However, the above LMXBs during quiescence were fainter than the detection limit of X-ray instruments, so their X-ray emission could have started rising well before the first X-ray detection (the derived X-ray delay could be much lower than reported). 

We monitored V404 with the 2-m FTN since 2006 and report here on the detailed 
analysis of the optical lightcurves from 2006 up to the time of the BAT trigger, 
and on the optical spectrum of V404 fortuitously collected with 
the William Herschel Telescope (WHT) $\sim13$ hours before the trigger.  
This is the closest in time an optical spectrum of an LMXB has been acquired 
prior to its first X-ray outburst detection. 
We discuss the results in the framework of the DIM (DHL).

\section{Observations and data reduction}
\label{sec:obs}

\subsection{Optical photometry}

Observations of V404 were taken with the 2-m Faulkes Telescope North (FTN, Haleakala, Maui, USA). Imaging was obtained in Bessell $V$, Bessell $R$ and Sloan Digital Sky Survey $i^{\prime}$ filters since April 2006, as part of a monitoring campaign of $\sim40$ LMXBs \citep[][]{lewis08}. We present more than nine years of data (from 2006 April 8 to 2015 June 9). Observations were typically made once per week when V404 was visible, and exposure times are 200 seconds in each filter. Automatic pipelines de-bias and flat-field the FTN science images.

There is a field star of magnitude $V=18.90\pm0.02$, $R=17.52\pm0.01$, $i^{\prime}=16.92\pm0.01$ just 1.4 arcsec north of V404 \citep{udalski91,casares93,iphas}. The two stars are blended in all images, so we performed aperture photometry (using \textsc{PHOT} in \textsc{IRAF}) adopting an optimum fixed aperture radius of 12 pixels (3.6'') to encompass the flux of both stars. The same aperture was used for photometry on four comparison stars 13--34 arcsec from V404. These were used for flux calibration, and themselves calibrated using field stars of known magnitudes listed in \cite{udalski91} for $V$-band, \cite{casares93} for $R$-band and the second data release of the IPHAS \citep[INT Photometric H$\alpha$ Survey of the Northern Galactic Plane catalogue;][]{iphas} for $i^{\prime}$. Magnitudes of V404 were obtained in total from 392 usable images.

\subsection{Optical spectroscopy}

V404 was observed on 2015 June 15 at UTC 0500 (MJD 57188.208) with the Intermediate dispersion Spectroscopic and Imaging System (ISIS) on the 4.2m WHT at Observatorio del Roque de Los Muchachos, La Palma, Spain. We obtained two 600 s exposures covering a total spectral range 4173--7153 \AA, utilising the R600B and R gratings and a 1" slit in photometric conditions of good ($\sim1$'') seeing. Spectra were reduced and extracted using standard IRAF procedures. The one-dimensional spectra were wavelength calibrated using a low-order polynomial fit to the pixel-wavelength arc data. We flux-calibrated the spectrum using the nearby flux standard star BD+25 4655 \citep{oke90}. 
The WHT observations were previously reported in \cite{munoz-darias15}.

\section{Results}
\label{sec:results}

\subsection{X-ray rise to outburst}

We convert the BAT 15-50 keV count rate averaged on each orbit \citep{krimm13}
at the time of trigger (MJD 57188.772) and the first $3\sigma$ upper 
limit before the trigger (MJD 57188.705) to unabsorbed 15-50 keV flux using \textsc{WebPIMMS}. 
Since the spectrum of the first flare is highly absorbed and the spectral shape not well constrained, 
we used a power law with slope $\Gamma=0.3-1.2$ \citep{kuulkers15}. 
We measure F$=4.0\pm0.5\times10^{-8}\rm\,erg\,cm^{-2}\,s^{-1}$ and F$<4.2\times10^{-9}\rm\,erg\,cm^{-2}\,s^{-1}$, respectively. V404 displays X-ray variability a factor of a few in quiescence \citep[][]{bernardini14}. 
We convert the lowest literature X-ray quiescent 0.3--10 keV flux \citep[2006 {\em XMM-Newton} pointing;][]{bradley07} to 15-50 keV flux using \textsc{WebPIMMS}, $\Gamma=1.85$ and N$_{H}=1\times10^{22}\rm\,cm^{-2}$ \citep{rana15}, and obtain F$\sim1\times10^{-12}$ $\rm\,erg\,cm^{-2}\,s^{-1}$. Assuming that the rise to outburst of the first X-ray flare was monotonic and slow, e.g. a single exponential rise 
starting from the quiescent level or above it, it must have begun after MJD 57188.530. 

The flares detected by {\em INTEGRAL} \citep{winkler03} at 25--100 keV 
after the BAT trigger show a fast rise ($\lesssim1$ h) and duration 
of hours \citep{rodriguez15b}. Moreover, during its 1989 outburst, 
the 1.2--37 keV lightcurve of V404 showed flares with very fast rises 
\citep[minutes;][]{kitamoto89,terada94,zycki99}. 
A complex behavior for the X-ray rise to outburst, 
e.g. a slow rise followed by a sudden increase, cannot be excluded since 
the source in quiescence is below the BAT detection limit.
We conclude that MJD 57188.53 can be safely considered as a lower 
limit to the rise to outburst of the first X-ray flare detected by BAT.

\subsection{Optical photometry}

\begin{figure*}
\begin{center}
\includegraphics[angle=0,width=5.5in]{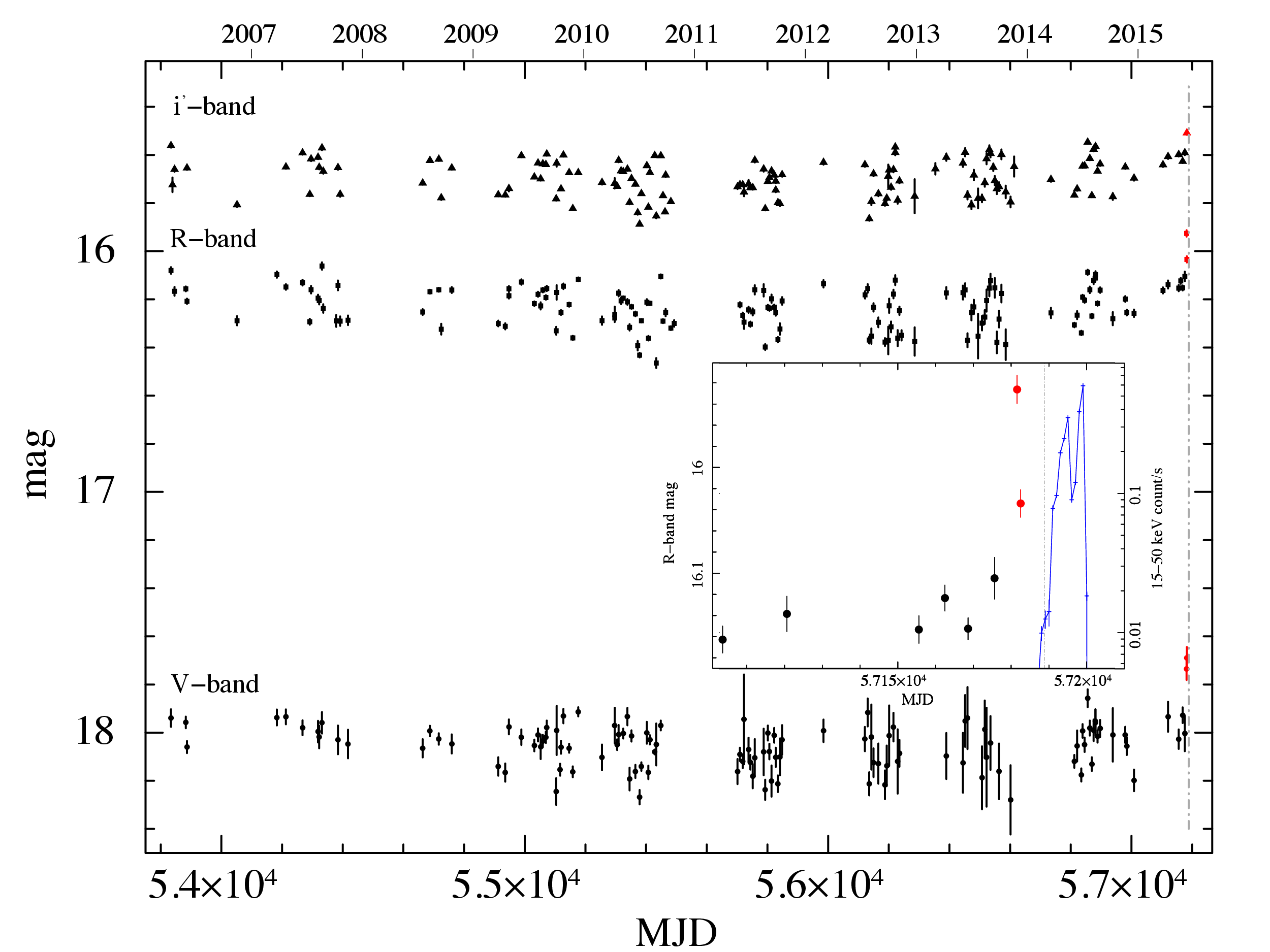} \\
\caption{Optical lightcurve in i$^{'}$ (triangles), R (squares), and V (circles) band from 2006 to 2015. Magnitudes include flux from the nearby contaminating star. The dot-dashed line at MJD 57188.772 shows the X-ray trigger. The red points are 2015 June 8 (MJD 57181.5) and 9 (MJD 57182.5). No observation in i$^{'}$ band was performed on 2015 June 8. In the insert, a zoom-in of the 2015 R band data compared with the $3\sigma$ BAT detections (blue crosses).}
\label{fig:lc}
\end{center}
\end{figure*}

\begin{figure*}
\begin{center}
\includegraphics[angle=0,width=5.4in]{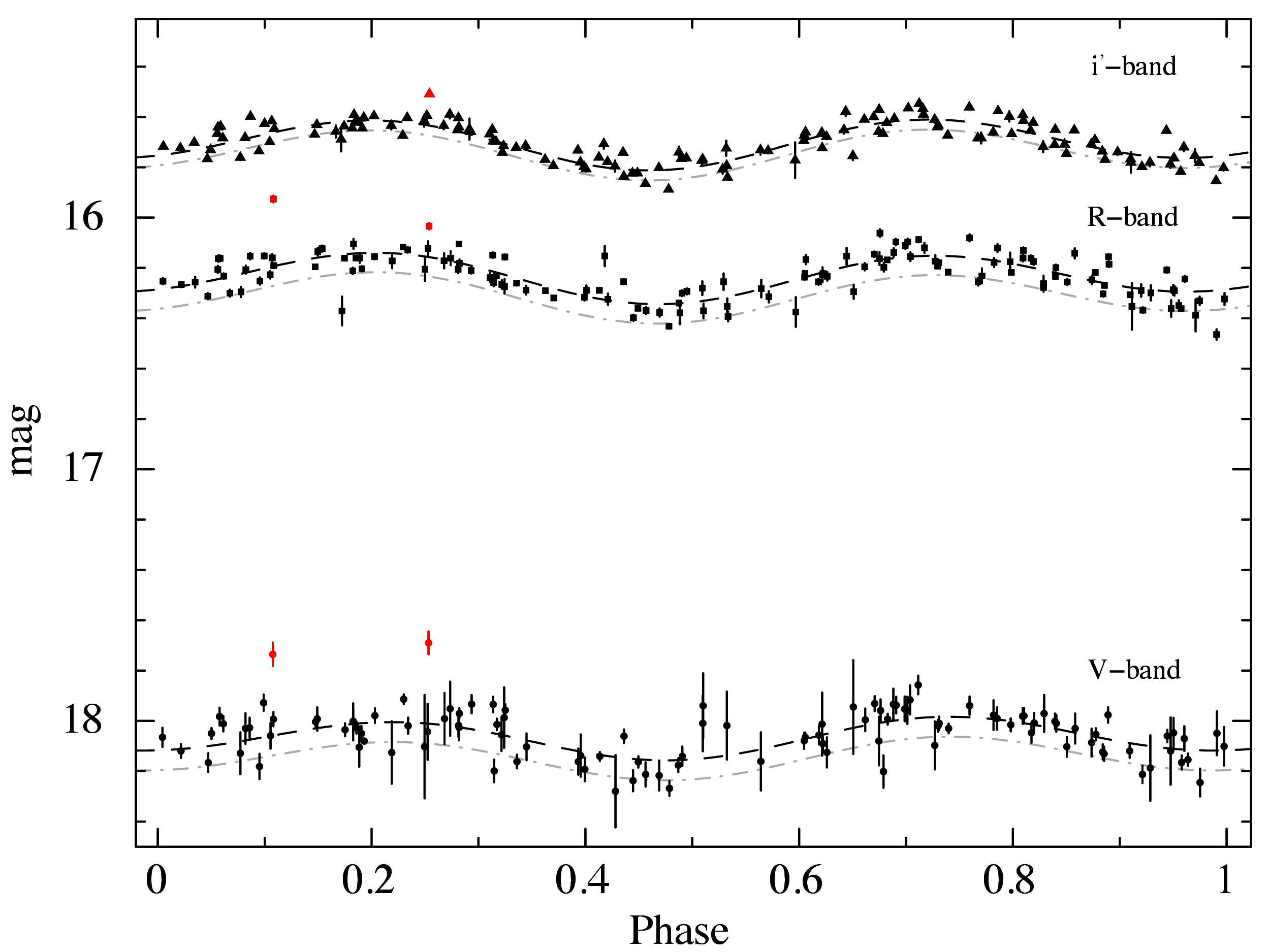} \\
\caption{Orbital lightcurves. The dashed line is a constant plus two sines best fitting model. The dot-dashed line is the lower envelope.}
\label{fig:mod}
\end{center}
\end{figure*}

In Fig. \ref{fig:lc} we show the optical lightcurve in V, R and i$^{'}$ bands, from 2006 up to the 2015 BAT trigger. A week before it (the last two points), the V, R and i$^{'}$ magnitudes are the highest recorded throughout this time. 

Using P$_{orb}=6.4714\pm0.0001$ day and T$_{0}=2448813.873\pm0.004\rm\,HJD$ as orbital ephemerides \citep{casares94}, we generate the orbital lightcurve (Fig. \ref{fig:mod}). 
The relative uncertainty on the phase determined with these ephemerides is $\sim0.05$, across the entire duration of our observations, but only $\sim0.005$ from one year to the next. The typical variation due to the tidally distorted donor's ellipsoidal modulation is present. Low amplitude flaring, likely due to residual accretion activity, is also present, as has been documented before \citep{shahbaz03,zurita04,hynes04,hynes09,bernardini14}.

\begin{figure*}
\begin{center}
\includegraphics[angle=-90,width=5.4in]{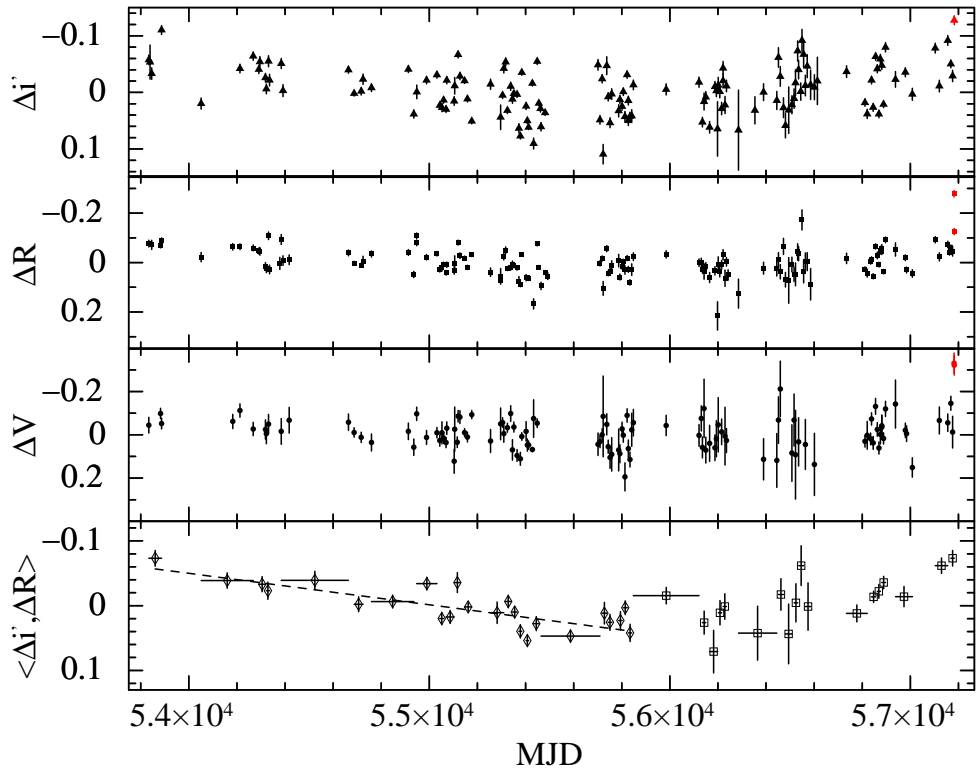} \\
\caption{Residual lightcurves. The bottom panel shows the average of the residual in the i$^{'}$ and R band. The dashed line is a fit made with a constant plus a linear component.}
\label{fig:res}
\end{center}
\end{figure*}

We fit the orbital lightcurves (except the last two points) with a constant plus double sinusoid function, whose phases are free to vary so as to account for unequal minima and maxima, as frequently seen in quiescent LMXBs including V404 \citep[][]{zurita04}. On 2015 June 8 (MJD$\sim57181.5$, $\phi\sim0.11$), and 9 (MJD$\sim57182.5$, $\phi\sim0.25$) the optical magnitude in all bands is significantly brighter than the average quiescent modulation level by 0.1--0.3 mag, and above the low amplitude flaring behavior (where $\Delta\,{\rm mag}\lesssim0.1$). We subtract the best fitting model from the orbital lightcurves, and show in Fig. \ref{fig:res} the residual lightcurves. We note that the last two points of the lightcurves have the highest residuals, that the residuals in the different bands look correlated on short timescales (days) and that a long-term trend (years) seems present (first a decay and then a rise). 
 
We measure the statistical significance of the correlation using Spearman's rank test on the R and i$^{'}$ bands, where the S/N is higher compared to V. The Spearman's coefficient is $\rho=0.78$, and the null hypothesis probability is $P=3.5\times10^{-27}$, so the residuals are positively correlated. This and the small error bars on each data point, suggest that the observed variability is intrinsic to the source and is likely accretion activity on timescales longer than minutes (the time between two consecutive exposures). 
We combine the i$^{'}$ and R band residuals. In the bottom panel of Fig. \ref{fig:res} we show the average of R and i$^{'}$ bands, $<\Delta\,i^{'},\,\Delta\,R>=0.5(<\Delta\,i^{'}>+<\Delta\,R>)$, where we use three points per bin. The decreasing-increasing trend is now clearer. We fit the first part of the lightcurve (up to MJD 55834.5) with a constant plus a linear function. An F-test gives a $4.4\sigma$ significance for the inclusion of the latter. Between MJD 53860 and 55834.5 we measure a decrease of $\sim0.02\rm\,mag/year$.

We detect an optical precursor to the first X-ray flare registered by BAT. Assuming that the two events are directly linked and that the rise of the X-ray flare is monotonic, we estimate a delay in the rise of the X-ray (MJD$\sim57188.5$) compared to the optical (MJD$\sim57181.5$) emission of at least 7 days. It could be as long as 13 days, considering that the optical flux may have started rising immediately after the pointing before the precursor (MJD$\sim57175.6$). The delay's length is similar to those seen in the other four LMXBs that showed indication of this behavior.
We have also found evidence of a long-term trend. The last 2 points of the residual lightcurve before the outburst are significantly above this quiescent downward trend. Consequently, we can constrain the long-term rise to have started before MJD 57129.5 (2015 April 17).
 
\subsection{Spectral Energy Distribution}

To construct the Spectral Energy Distribution (SED) of the optical precursor, we first measure the lower envelope of the modulation \citep[see][]{zurita04}, which is the donor contribution at each phase (Fig. \ref{fig:mod}). We combine the 2015 June 8 and 9 images (in the i$^{'}$ band only June 9 is available), and we subtract the lower envelope from the derived magnitudes.
We de-redden the fluxes of the residuals using A$_{V}=4$ \citep{casares93,hynes09} and the extinction law of \cite{cardelli89}, and measure a precursor spectral index $\alpha=-0.35\pm0.32$, where $F_{\nu}\propto\,\nu^{\alpha}$.  
This is consistent with a $7500\pm1500$ K blackbody, which peaks in the visible. These errors do not take into account any uncertainties in the extinction A$_{V}$ \citep[][]{hynes09}.

The SED could be consistent, within $1\sigma$ confidence level, with both optically-thin synchrotron emission with $\alpha\sim-0.7$, or optically-thick (flat) synchrotron emission with $\alpha\sim0$. However, Bernardini et al. (2016, submitted) showed that during the 1989 outburst the jet dominates the optical flux in the hard state, but makes marginal contribution in quiescence. 

We measured a de-reddened spectral index from the WHT spectrum of $\alpha=-1.85\pm0.08$ for the continuum, by removing the hydrogen lines from the red arm, similar to the spectral index of some flares seen during quiescence \citep{shahbaz03}; this suggests a variable spectrum during the initial rise into outburst.

\subsection{Optical spectroscopy}

\begin{figure}
\begin{center}
\includegraphics[angle=0,width=3.66in]{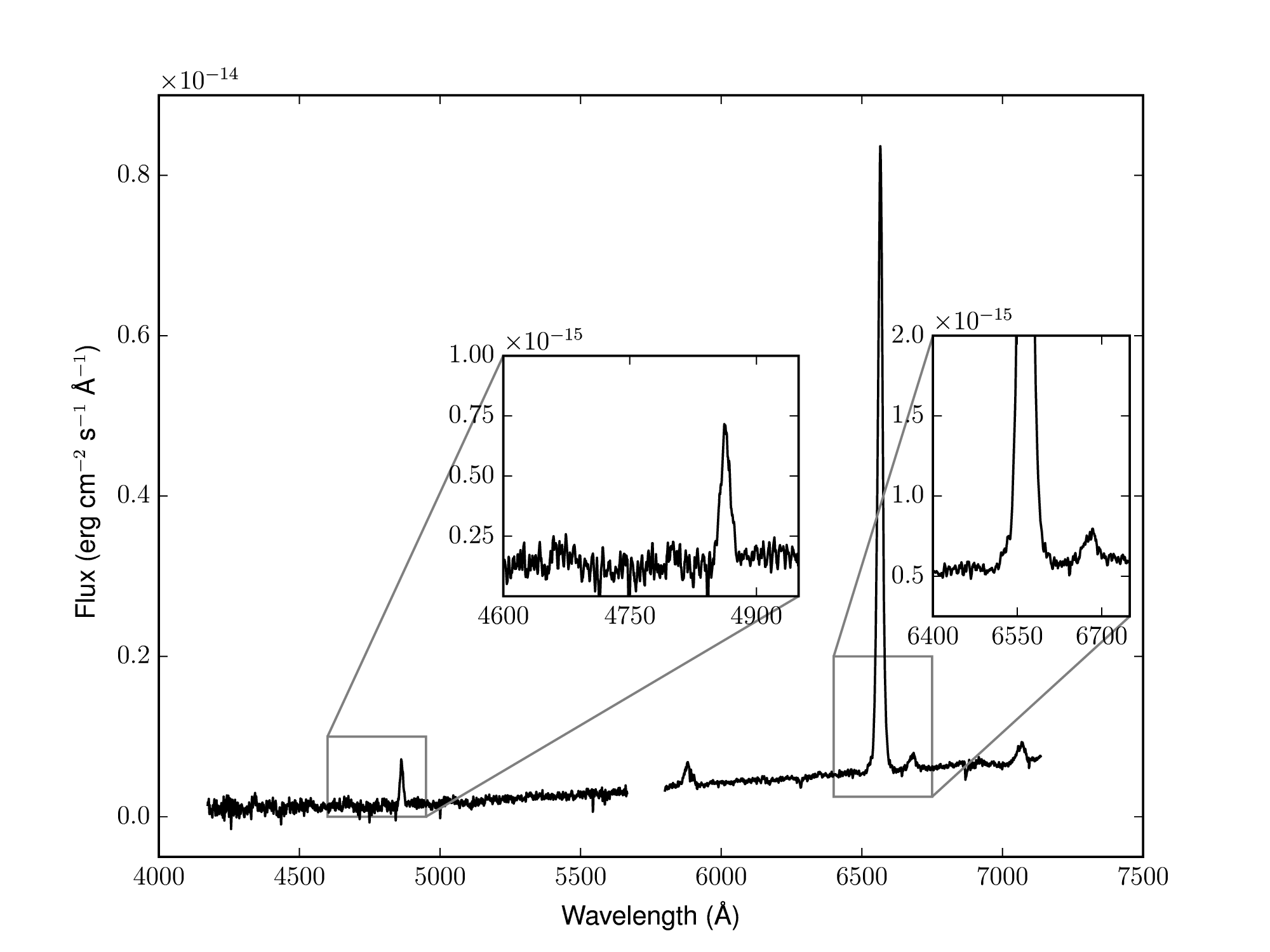} \\
\caption{Averaged optical spectrum (smoothed using a 5-point boxcar algorithm) obtained with WHT/ISIS on 2015 June 15. The spectrum has been flux calibrated using the flux standard star BD+25 4655. The left and right insets show the zoomed in region around H$\beta$ and the base of H$\alpha$, respectively.}
\label{fig:spec}
\end{center}
\end{figure}

\begin{figure}
\begin{center}
\includegraphics[angle=0,width=3.2in]{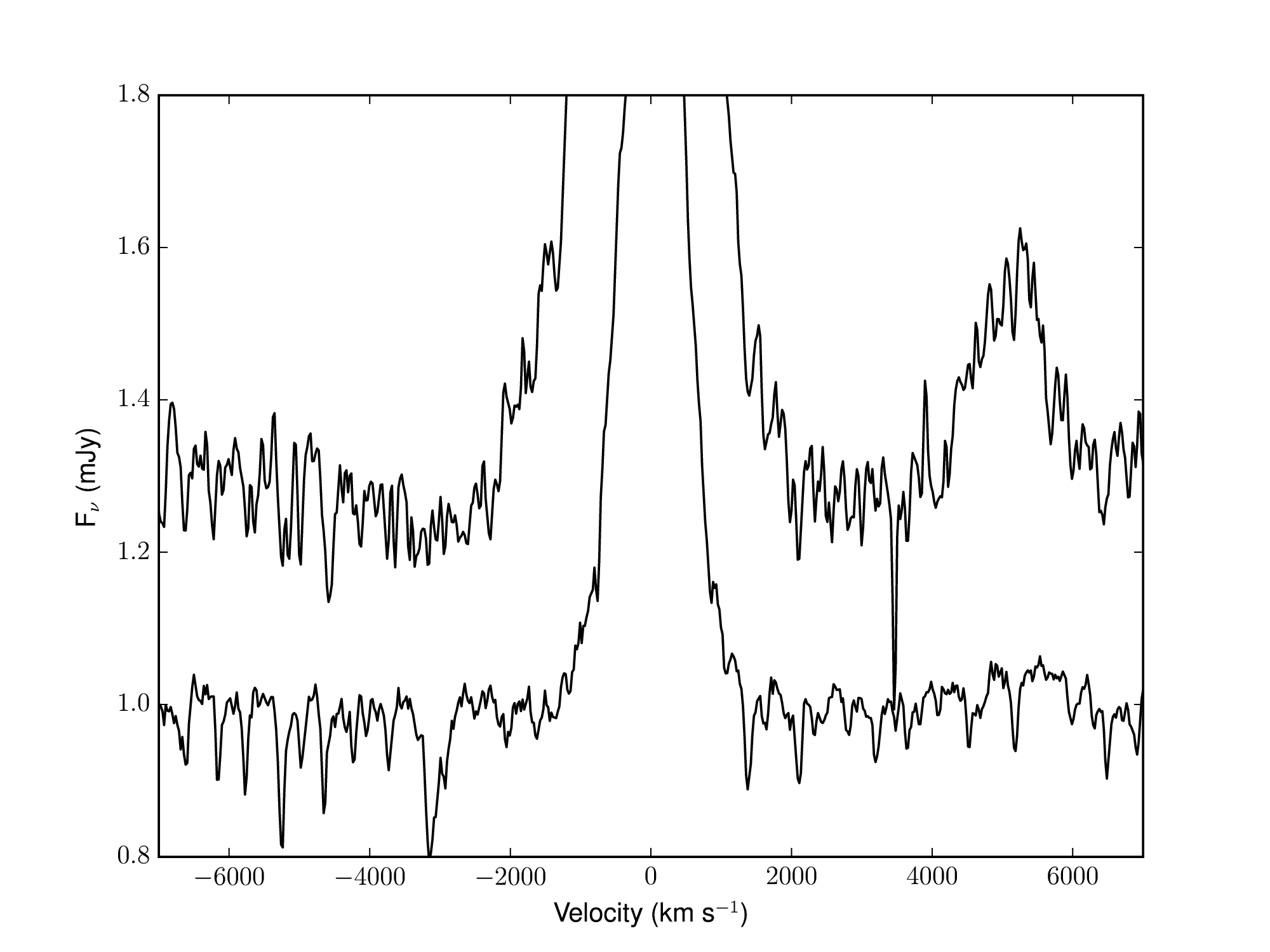} 
\caption{June 15 pre-outburst spectrum (offset by 0.3 mJy and smoothed with a Gaussian kernel of FWHM$=$2 pixels) compared with the 20 year average quiescent spectrum (bottom), zoomed in on the region surrounding H$\alpha$. Spectra have been averaged in the rest frame of the companion.}
\label{fig:halpha}
\end{center}
\end{figure}

The optical spectrum is dominated by a strong H$\alpha$ emission line (Fig. \ref{fig:spec}). H$\beta$ is also present, along with multiple He I emission lines. However, He II (4686 \AA), typical of X-ray illuminated disks in outburst, is absent. At first sight the absorption features of the companion seem also absent, but a closer look reveals clear identifications hidden within the noise (Fig. \ref{fig:halpha}). A cross-correlation of the H$\alpha$ pre-outburst spectrum and the average of 220 quiescent spectra of V404 obtained between 1990 and 2009 \citep[see][]{casares15} with the spectrum of the radial velocity template HR 8857 yields clear peaks at heliocentric velocities consistent with those of the donor star at the orbital phase of our spectra. 
The $\sim7\times$ increase in H$\alpha$ flux compared with the quiescent level \citep{casares92b} suggests that the accretion disk is much brighter than in quiescence.  
\indent We estimate the inner radius of the truncated disk, $R_{in}=0.5(c\,sin(i)/v_{in})^{2}$, by studying H$\alpha$ in emission. By measuring its half-width at zero intensity (HWZI) we estimate the velocity at the inner edge of the disk $v_{in}$ \citep[][]{narayan96}. We use a non-linear least squares algorithm to fit a low-order polynomial to the continuum (masking H$\alpha$) and subtracting it. We apply the same approach to fit a double Gaussian to the H$\alpha$ profile, finding it more accurate than a single Gaussian due to a broader component being present at the base of the line (Fig. \ref{fig:spec}). We utilize the lower amplitude broader Gaussian to estimate the HWZI and hence $v_{in}$ by measuring 5$\sigma$. \\
\indent From fits to the lower amplitude broad Gaussian we find a slightly redshifted H$\alpha$ peak at $6564.3\pm0.1$ \AA\ with HWZI$=54.0\pm0.5$ \AA, which translates to $v_{in}=2468\pm23\rm\,km\,s^{-1}$. This is actually a lower limit on $v_{in}$, as higher velocity structure is present around the base of the line profile, though difficult to fit. Using the derived inclination of $i=67^{\circ}$ \citep{khargharia10} we find that $R_{in}<6200$ Schwarzschild radii ($R_{s}=2GM_{BH}/c^{2}$) at MJD 57188.208. 

For comparison, the HWZI of the average quiescence H$\alpha$ line has HWZI$\lesssim1500\rm\,km/s$, implying $R_{in}\gtrsim17000\rm\,R_s$. Therefore, the inner disk radius may have decreased by a factor $\sim3$ in our pre-outburst spectrum with respect to quiescence. 

\section{Discussion}
\label{sec:disc}


The DIM predicts \citep[see Eq. 51 in][]{lasota01} that in LMXBs, only inside-out outbursts can occur (e.g. the instability is triggered well inside the accretion disk and will propagate outwards). 
Inside-out outbursts do not start exactly at the inner disk edge and fronts propagate both ways. To explain the long recurrence time and the intensity of the outbursts of LMXBs, and their quiescent X-ray luminosities, the DIM requires that the inner quiescent disk is truncated far from the compact object (DHL). The heating fronts propagate rapidly with a speed $\sim\alpha\,c_s$, where $\alpha$ is the the hot disk viscosity parameter and $c_s$ the speed of sound. The in-going front dies out quickly reaching the truncation radius without affecting much of the disk structure and then the hole is refilled in a viscous time. Since the disk X-ray emission is mainly emitted in the disk inner regions, a delay of several days in the rise to outburst of the X-ray with respect to the optical emission is expected. In the case of a non-truncated disk the delay would be at most 1 day (DHL).

A description of the rise to outburst can be found in Sect. 5.1 of DHL. We discuss our results according to this latest version of the DIM. The delay represents the difference between the time $t_V$, when the outburst starts at some $R(V)$ of the truncated disk and the time $t_X$, when the moving-in inner disk edge reaches the radius $R(X)$, where $R(V)>R(X)$, attaining the temperature that allows emission of X-rays. 
The X-ray delay ($\Delta\,t_{\rm\,V-X}$) corresponds to the difference between the two radii.
From Eq. 16 of DHL we can estimate $R(V)$ if we assume that the 7 days X-ray delay corresponds to the viscous time ($t_{\rm vis}=\Delta t_{\rm V-X}$). We get $\Delta\,t_{\rm\,V-X}=15.3\,M_{10}^{1/2}\alpha_{0.2}^{-1}T_5^{-1}\left(R_{10}^{1/2}(V)-R_{10}^{1/2}(X)\right)$ days, where $M_{10}=M_{BH}/10\,M_{\odot}$, $\alpha_{0.2}=\alpha/0.2$ is the hot branch viscosity parameter, $R_{10}=(R/10^{10}\rm\,cm)$ is the disk radius, 
and $T_{5}=(T/10^{5}\rm\,K)$ is the midplane temperature, where $T\gtrsim3-4\times10^{4}$ K 
at the start of the outburst \citep[see][]{lasota08}. We use $T=30000-50000\rm\,K$, $\alpha=0.1-0.2$, and $R(X)=5\times10^{8}$ cm (as in DHL), and we derive that the instability corresponding to the outburst precursor (MJD 57181.5) may have been triggered at $R(V)\sim0.9-2.2\times10^{9}\rm\,cm$  ($\sim340-830\,R_{s}$). We notice that the disk size is $\sim10^{12}\rm\,cm$

We also derive a direct constraint on $R(V)$ from the HWZI of H$\alpha$. 
At MJD 57188.208, 13 hours before the first X-ray detection ($\sim6.5$ days after the optical precursor detection), $R(V)<6200\,R_{s}$. This is an upper limit to $R(X)$, since $R(X)<R(V)$, and so it represents a constraint on the size of the inner edge of the truncated disk ($R_{in}$), close to the X-ray outburst onset. We notice that this upper limit is consistent with $R(V)$ derived from DIM equations.

We can use Eq. A.1 in \cite{lasota08} to estimate the critical effective disk temperature ($T^{+}_{eff}$) needed to ionize hydrogen and start the outburst. Using the derived $R_{in}$, $T^{+}_{eff}\approx7000\rm\,K$ (the dependency on R, and M in the equation is weak). We note that at the time of the optical precursor the disk temperature ($7500\pm1500\rm\,K$) is consistent with $T^{+}_{eff}$.

All versions of the DIM predict a constantly increasing optical flux during quiescence \citep[][]{lasota01}, while observed quiescent fluxes of dwarf novae and LMXBs are constant or decreasing \citep[however, see][]{wu15}.
We observe a long-term trend in the optical magnitude of V404, a 0.1 mag decrease followed by a 0.1 mag increase. 
The source is known to show year-to-year optical modulation changes of similar intensity \citep[see Fig. 1 in][]{zurita04}. Most of the variability we detect above the orbital modulation is likely due to accretion activity and occurs at all orbital phases (Fig. \ref{fig:mod}).

It is known that accretion is happening at a low level in quiescence (short term variability has been seen in X-ray, optical and radio). Small accretion rate changes from year to year, likely explain the long-term optical variations.
The rise in the optical flux from April 2015 may reflect a recent increase in $\dot{M}$ eventually culminating in the outburst. The disk becomes progressively hotter as matter builds up and when the temperature reaches the ionization level, the inside-out heating wave quickly propagates from the trigger site, close to the inner edge of the truncated disk, through the whole disk (in both directions). 
Then, the inner edge of the truncated disk moves inwards, on the longer viscous timescale, and a week after the optical precursor, when $R_{in}<6200\,R_{s}$ (a factor of $\sim3$ lower than in quiescence), it is hot enough to generate X-ray emission and we observe the first X-ray flare. 

\section*{Acknowledgments}

The FTN is maintained and operated by Las Cumbres Observatory Global Telescope Network. 
JPL was supported by the CNES. JC acknowledges support by MINECO under grants AYA2013-42627 and PR2015-00397

\bibliographystyle{mn2e}
\bibliography{biblio}

\vfill\eject
\end{document}